\documentclass[]{aa}
\usepackage{amsmath}
\usepackage{latexsym}
\usepackage{longtable}
\usepackage{natbib}
\usepackage[latin1]{inputenc}
\usepackage{graphics}
\usepackage{epsfig}
\usepackage{graphicx}

\begin{document}

\title{Kolmogorov stochasticity parameter measuring the randomness in Cosmic Microwave Background}
\author{V.G.Gurzadyan\inst{1},  A.A.~Kocharyan\inst{1,2}}

\institute
{\inst{1} Yerevan Physics Institute and Yerevan State University, Yerevan,
Armenia\\
\inst{2} School of Mathematical Sciences, Monash University, Clayton, Australia
}

\date{Received (\today)}

\titlerunning{CMB maps}

\authorrunning{V.G.Gurzadyan and A.A.Kocharyan}

\abstract{
The Kolmogorov stochasticity parameter (KSP) is applied to quantify the degree of randomness (stochasticity) in the temperature maps of the Cosmic Microwave Background  radiation maps. It is shown that, the KSP for the WMAP5 maps is about twice higher than that of the simulated maps for the concordance $\Lambda$CDM cosmological model, implying the existence of a randomizing effect not taken into account in the model. As was revealed recently, underdense regions in the large scale matter distributions, i.e the voids, possess hyperbolic and hence randomizing properties. The degree of randomness for the Cold Spot appears to be about twice higher than the average over the mean temperature level spots in the sky, which supports the void nature of the Cold Spot. Kolmogorov's parameter then acts as a quantitative tracer of the voids via CMB.}


\maketitle

\section{Introduction}

The statistic introduced by Kolmogorov in 1933 \cite{K} is based on a stochasticity parameter of a given sequence of real numbers. This measurement of stochasiticity is successfully applied to measure the objective randomness degree of finite sequences resulting due to dynamical systems or in number theory \cite{A_KSP}.

Here we apply the KSP in a physical problem of cosmological interest, namely, to the properties of the Cosmic Microwave Background (CMB) radiation which is one of the basic sources of information on the early Universe and its present structure. Recent studies include non-Guassianities reported for the CMB properties, such as the multipole alignments, hemisphere anomalies, temperature independent ellipticity of the excursion sets, large-scale symmetry, see (Eriksen et al 2007, Copi et al 2007, Gurzadyan \& Torres 1997, Gurzadyan et al 2005, 2007, 2008). Among the signals having low probability for the case of Gaussian fluctuations is a region in the southern sky known as the Cold Spot \cite{c_spot,Cruz}. 

The present study concerns the CMB non-Gaussianties and also to the voids, i.e. underdense regions in the large scale matter distribution. The idea is as follows. On the one hand, the voids are one of the discussed phenomena associated to CMB, particularly as responsible for the nature of the Cold Spot anomaly, see \cite{CMB_void,CardS,Uzan,Das}. On the other hand, as it was shown recently for the Friedmann-Robertson-Walker universe with perturbations of the metric, the voids can act as hyperbolic lenses \cite{GK1}; similar conclusion on the voids acting as divergent lenses is drawn in \cite{Das}. As follows from the theory of dynamical systems, the hyperbolicity of the geodesic flows, Anosov flows particularly known in this relation, implies the decay of time correlation functions and a loss of information on the initial conditions \cite{Arnold}. 

So, if the voids are hyperbolicity regions, they are able to randomize the distribution of the null geodesics and hence the temperature distribution in the CMB maps. Then, the line-of-sight distribution of the voids have to be of particular importance in the degree of randomization. Would it be due to a void, the case of the Cold Spot can be indicative.      

To reveal this effect of randomization we estimate the KSP both for the CMB temperature map and for the map simulated for the power spectrum defined parameters of the $\Lambda$CDM concordance model. We then estimate the KSP for the Cold Spot and for threshold temperature regions of the same size in the observed temperature map. Thus the randomization along with the ellipticity in CMB maps, can be an observational tracer for the hyperbolicity.

\section{Kolmogorov Stochasticity Parameter and\\ Statistic}

Let $\{X_1,X_2,\dots,X_n\}$ be $n$ independent values of the same real-valued random variable $X$ ordered in an increasing manner $X_1\le X_2\le\dots\le X_n$ and let
$$
F(x) = P\{X\le x\}\ 
$$
be a cumulative distribution function (CDF) of $X$. Their {\it empirical distribution function} $F_n(x)$ is defined by the relations
\begin{eqnarray*}
F_n(x)=
\begin{cases}
0\ , & x<X_1\ ;\\
k/n\ , & X_k\le x<X_{k+1},\ \ k=1,2,\dots,n-1\ ;\\
1\ , & X_n\le x\ .
\end{cases}
\end{eqnarray*}

Kolmogorov's stochasticity parameter $\lambda_n$ has the following form  \cite{K, A_KSP}
\begin{equation}\label{KSP}
\lambda_n=\sqrt{n}\ \sup_x|F_n(x)-F(x)|\ .
\end{equation}

Kolmogorov proved in \cite{K} that for any continuous CDF $F$
$$
\lim_{n\to\infty}P\{\lambda_n\le\lambda\}=\Phi(\lambda)\ ,
$$
where $\Phi(0)=0$,
\begin{equation}
\Phi(\lambda)=\sum_{k=-\infty}^{+\infty}\ (-1)^k\ e^{-2k^2\lambda^2}\ ,\ \  \lambda>0\ ,\label{Phi}
\end{equation}
the convergence is uniform, and $\Phi$ (Kolmogorov's distribution) is independent on $F$.
KSP is applied to measure the objective stochasticity degree of datasets \cite{A_KSP}. It is easy to see that the inequalities $0.3\le\lambda_n\le 2.4$ can be considered as practically certain \cite{K}.

\section{Analysis}

The temperature maps of the CMB are sequences of numbers indicating the temperature assigned to the pixels in certain coordinate representation, commonly in HEALPix \cite{HP}.

For the analysis we used the latest, 5-year data obtained by the Wilkinson Microwave Anisotropy Probe (WMAP) at 94 GHz (3.2mm, W-band). This band has the best resolution among the other bands and is the least contaminated one by the Galactic synchrotron radiation. 

The Cold Spot region is assigned here as an area of 3$^{\circ}$ radius centred in $l$=208$^{\circ}$.7, $b$=-55$^{\circ}$.6,  represented as a string of temperature values of 2155 pixels varying within the interval $T_{min}= -445.6$ and $T_{max}= 263.0$ (in $\mu$K).    

The KSP has been calculated for 150 regions of the size of the Cold Spot but neither cold or hot, i.e. with $\bar{T}\simeq 0$, randomly distributed over the sky, excluding, as usual, the Galactic disk region $|b|< 20^{\circ}$, as well as for 150 simulated regions of the same size and condition. The simulations have been performed via the standard scheme using the HEALPix package ({\it synfast}), superposing the noise of WMAP with its FWHM beam smoothing (see e.g. Gurzadyan et al 2005, 2007, 2008).  

For each dataset $k$ we assume that $F(x)$ is CDF for Gaussian distribution, calculate $\lambda_n^{(k)}$ in Eq.(1), estimate $\Phi(\lambda_n^{(k)})$ in Eq.(2), and finally get the mean probability $\Phi$.

The results for the mean values and variances for the observed WMAP's map, of the simulated map, and for the Cold Spot are given in the following Table: 
\begin{table}[ht]
\centering
\begin{tabular}{l c c}
\hline\hline
Source                  &  Mean($\Phi$) & Var($\Phi$)\\ [0.5ex]
\hline
WMAP's data             & 0.353  & 0.07 \\
Simulations             & 0.223  & 0.04 \\
Cold Spot (W band)      & 0.749  & 0.00 \\
Cold Spot (FR:W,Q,V bands) & 0.859  & 0.01 \\ [1ex] 
\hline
\end{tabular}
\end{table}

Note, to trace the role of the dust, KSP for the Cold Spot was estimated also for the WMAP's foreground reduced maps at W, Q and V bands.    

\section{Conclusions}

The results indicate that the real CMB sky is random with twice higher degree than the simulated one, i.e. there is an extra randomizing effect not included in the cosmological model. Respectively, the Cold Spot is about twice more random than similar regions over the sky. Would this be due to the randomizing effect of the hyperbolic voids, it can particularly support the void nature of the Cold Spot. Kolmogorov's parameter then can serve as a tool to probe the voids in the Universe via the properties of CMB. Thus, this study has to be considered as a preliminary one to be followed with detailed analysis of the regions of different degree of randomness in CMB maps.       

The considered application of Kolmogorov's parameter can justify its further use in physical problems, since in spite of an apparent abstract content, it appears calculable and insightful.     

We are thankful to the referee for valuable comments and to A.Kashin for discussions and help.

\end{document}